\documentclass[dvips]{article}

\usepackage{icrc2011}
\usepackage{amsmath}
\usepackage{units}

\title{Very-high-energy gamma radiation from supernova remnants as seen with H.E.S.S.}

\newcommand{\g}{$\gamma$}
\newcommand{\gr}{$\gamma$-ray}
\newcommand{\grs}{$\gamma$ rays}
\newcommand{\etal}{\MakeLowercase{\textit{et al.}}} 
\shorttitle{Anne Bochow \etal Very-high-energy \g\ radiation from supernova remnants as seen with H.E.S.S.}

\authors{Anne Bochow$^{1}$, Svenja Carrigan$^{1}$, Henning Gast$^{1}$, Vincent Marandon$^{1}$, Matthieu Renaud$^{2}$, Werner Hofmann$^{1}$ for the H.E.S.S. collaboration}
\afiliations{$^1$Max-Planck-Institut f\"ur Kernphysik, , P.O. Box 103980, D 69029 Heidelberg, Germany\\
$^2$Laboratoire de Physique Th\'{e}orique et Astroparticules, Universit\'{e} Montpellier 2, CNRS/IN2P3, CC 70, Place Eug\`{e}ne Bataillon, F-34095 Montpellier Cedex 5, France}
\email{anne.bochow@mpi-hd.mpg.de}

\abstract{Very-high-energy (VHE, E $>$ 100 GeV) \g\ radiation has already been detected 
from several supernova remnants (SNRs). These objects, which are well-studied in 
radio, optical and X-ray wavelengths, constitute one of the most intriguing source 
classes in VHE astronomy. H.E.S.S., an array of four imaging atmospheric Cherenkov 
telescopes in Namibia, has recorded an extensive dataset of VHE \gr\ 
observations covering the central region of the Milky Way, both from pointed 
observations as well as from the Galactic Plane Survey conducted in the inner region 
of the Galaxy. From radio observations, several hundred SNRs are known in the Milky 
Way, but until now only few of them have been identified as VHE \gr\
emitters. Using the H.E.S.S. dataset and a large ensemble of radio SNRs localized in 
the inner region of the Galaxy, the standard framework that links the origin of 
cosmic rays to the \gr\ visibility of SNRs can now be tested. Here we present 
the ensemble of investigated SNRs and discuss constraints on the parameter space 
used within a theoretical model of hadronic VHE \gr\ production.}
\keywords{ VHE gamma radiation -- SNR -- Galactic }

\begin{document}
\maketitle
\section{Introduction}
SNRs are widely believed to be the main sources of Galactic Cosmic Rays (CRs) with energies of up to \unit[$10^{15}$]{eV}. The amount of energy released during a supernova (SN) explosion, the rate of Galactic SNe and the energy density of CRs together with diffusive shock acceleration as a possible acceleration mechanism make SNRs the prime candidates for the acceleration of CRs. In radio, optical and X-ray wavelengths deep observations on SNRs have been performed and several SNRs have been discovered in the relatively young field of VHE \gr\ astronomy . With H.E.S.S. the shell-type morphology of individual SNRs such as RX~J1713.7-3946~\cite{HESS:1713I}, SN~1006~\cite{HESS:SN1006}, Vela Junior~\cite{HESS:VelaJunior} and HESS~J1731-347~\cite{HESS:1731} has been resolved. These discoveries support the hypothesis of particle acceleration in the shell of SNRs. Using the large dataset of H.E.S.S. accumulated during the Galactic plane survey (GPS)~\cite{ICRC2011:Survey} in recent years we study SNRs in the inner Galactic region, ranging from 275$^\circ$ to 60$^\circ$ in Galactic longitude and from -3$^\circ$ to 3$^\circ$ in Galactic latitude.

Here, the VHE \gr\ signal of known radio SNRs listed in Green's radio SNR catalog~\cite{GREEN:RadioSNRs} is investigated. For each SNR within the GPS region, an upper limit on the VHE \gr\ flux has been calculated and can be compared to the model predictions. In this work we test the commonly used model of the hadronic \gr\ emission expected from SNRs presented by Drury~\etal~\cite{Drury:SNRVisibility}. They calculate the integral \gr\ flux from hadronic interactions based on intrinsic SN parameters. VHE \g\ rays result from the decay of neutral pions generated in proton--proton interactions in the expanding shell of the SNR. For the first time, the high sensitivity of H.E.S.S. and the large dataset of the GPS allow us to derive upper limits that are low enough to test model parameters. Using the upper limit on the VHE \gr\ flux represents a conservative test since an additional leptonic component might be present in the flux.

\section{Probing the hadronic model}
Assuming that CRs are accelerated and VHE \grs\ produced in SNRs, the flux of VHE \grs\ depends on explosion and acceleration parameters, the properties of the ambient medium, and the distance of the SNR. Drury~\etal~\cite{Drury:SNRVisibility} calculate the expected integral flux of VHE \grs\ from SNRs generated from hadronic CR interactions above a given energy threshold \emph{E}:
\begin{eqnarray}
F_\gamma (> E) & \approx &  9 \cdot 10^{-11} \theta  \left( \frac{E}{\unit[1]{TeV}} \right) ^{-1.1}  \left( \frac{E_\mathrm{SN}}{10^{51} \mathrm{erg}} \right) \nonumber \\
               &         &   \left( \frac{d}{\unit[1]{kpc}} \right)^{-2}  \left( \frac{n}{\unit[1]{cm^{-3}}} \right)    \mathrm{cm^{-2}s^{-1}}, 
\label{eqn:fluxprediction}
\end{eqnarray}
where \emph{$\theta$} is the efficiency of the particle acceleration, \emph{E$_\mathrm{SN}$} the total energy released during the supernova explosion, \emph{d} the distance to the SNR and \emph{n} the density of the interstellar medium surrounding the SNR. Typical values are $\theta  =  0.1$, $E_\mathrm{SN}  =  10^{51}\mathrm{\,erg}$ and $n =  1\mathrm{\,cm^{-3}}$~\cite{Drury:SNRVisibility}. For this model the authors assume a power law with a spectral index of $\Gamma=2.1$. 

\section{SNR sample}
203 SNRs listed in Green's radio SNR catalog are located within the H.E.S.S. GPS region. For these SNRs an upper limit on the VHE \gr\ flux can be derived based on H.E.S.S. observations. A subsample used for the investigation of the hadronic model is chosen following theoretical considerations. The essential criterion for VHE \gr\ production is the evolutionary stage of the respective SNR. Except for the very young SNR \emph{G1.9+0.3} with an age of \unit[$\sim$110]{years}~\cite{Carlton:G0019+003}, all SNRs with known age can be expected to have reached the evolutionary stage of the Sedov phase where efficient particle acceleration takes place. An upper age bound for the VHE \gr\ visibility of SNRs has been discussed in several publications. Drury~\etal~\cite{Drury:SNRVisibility} give an age of \unit[$\sim$10$^4$]{years} where temperatures drop below \unit[$\sim$10$^6$]{K} and energy losses due to thermal cooling and particle escape from the shell due to magnetic instabilities become more important. Ptuskin~\&~Zirakashvili~\cite{Ptuskin:SNRdevelopment} study nonlinear effects in SNRs leading to a decrease of the maximum energy of accelerated particles. A generic prediction of the models they consider is a reduced VHE \gr\ visibility of SNRs at an age larger than a few thousand years. In the present work we therefore restrict the investigation of the hadronic model to SNRs with a known age of up to \unit[5000]{years}. 

Predicting the VHE \gr\ flux of SNRs requires a reliable estimate of their distance. These distance estimates are usually obtained through the comparison of HI absorption spectra with respective HI and molecular ($^{12}$CO, $^{13}$CO) emission spectra along the line of sight (see \cite{Tian:SNRdistance} and references therein). The uncertainties of the Galactic rotation curve model used to estimate these kinematic distances, together with those on the distance to the Galactic Center and the circular velocity of the Sun lead to errors of the order of 10-30\%. For relatively nearby (a few kpc away) SNRs with detected radio, optical or X-ray emission, it is sometimes possible to measure the expansion rate of the shell (or otherwise, that of the associated pulsar wind nebula, see \cite{Bietenholz:SNRdistance} for G21.5-0.9) which constrains the dynamical age (subject to the evolutionary stage of the SNR) and, therefore, provides information about the distance to the source (see \cite{Katsuda:SNRdistance} for Vela Junior).

The 20 SNRs considered within this study are listed in Table~\ref{table_SNRs} together with their mean radius $R$, distance $d$, age, the VHE \gr\ flux upper limit $(\Phi^{UL}, \mathrm{sec.}~\ref{sec:fluxUL})$ and the derived upper limit on the model parameters $(\theta E_\mathrm{SN} n^{UL}, \mathrm{sec.}~\ref{sec:paramUL})$. 
The mean radius $R$ is derived from the major and minor diameter given in Green's catalog as the geometric mean $R=\nicefrac{1}{2}\sqrt{{D_\mathrm{maj}}\cdot {D_\mathrm{min}}}$. The SNR positions have also been taken from the catalog. For the analysis the same spectral index of $\Gamma = 2.1$ has been adopted from the model of Drury~\etal\ for all SNRs, allowing us to perform a standardized analysis of the entire SNR sample. For some sources the spectral index is known to be other than 2.1 which leads to differing analysis results compared to a detailed spectral analysis~(see Aharonian~\etal\ \cite{Aharonian:G0009+001} for G0.9+0.1, HESS~J1747-281).

\section{Upper limit on the VHE \gr\ flux}
\label{sec:fluxUL}
For the calculation of a flux upper limit, a reference flux with a predefined
spectral shape has to be assumed,
\begin{equation}
\label{eq:refflux}
\phi_\mathrm{ref}(E)=\phi_0\,(E/E_0)^{-\Gamma}
\end{equation}
The normalization can be chosen arbitrarily as it will cancel out in
the end. We use $E_0=1\,\mathrm{TeV}$ in the following.  
For any given run with live-time
$T$, the number of expected signal events from a source can then be calculated as
\begin{equation}
\label{eq:expcounts}
N_\mathrm{run}=T\,\int_{E_\mathrm{min}}^\infty \phi_\mathrm{ref}(E)\,A_\mathrm{eff}(E,q)\,\mathrm{d}E
\end{equation}
Here, $E$ is energy,
$A_\mathrm{eff}$ is the effective area, $q$ symbolizes the observation
parameters (e.g.~the zenith angle) and $E_\mathrm{min}$ is the
threshold energy appropriate for the run. The individual expected
counts can then be summed over all runs to yield the total number $N$
of expected counts.

An upper limit on the integral flux above energy $E_1$ is then
calculated as
\begin{equation}
\label{eq:ul}
\Phi^\mathrm{UL}=\frac{n^\mathrm{UL}}{N}\,\int_{E_1}^{\infty}\phi_\mathrm{ref}(E)\,\mathrm{d}E
\end{equation}
where $n^\mathrm{UL}$ is the upper limit on the excess counts above
threshold that can be obtained from the measured number of photon
candidates above threshold within the signal region, $N_\mathrm{ON}$,
the corresponding number of counts from a suitable background control
region, $N_\mathrm{OFF}$, and the background normalization factor
$\alpha$~\cite{ref:berge}, by using the profile
likelihood method of Rolke \etal~\cite{ref:rolke}. 

The signal region is taken to be a circle of radius $\theta$, where
$\theta=0.1^\circ$ for SNRs with radius $R\leq{}0.05^\circ$ and $\theta= \sqrt{R^2 + 3 \cdot {\theta_\mathrm{PSF}}^2}$
for SNRs with radius $R>0.05^\circ$. This means that small SNRs are
treated as pointlike sources and $\theta$ is chosen such that the
significance gain $\sigma/\sqrt{t}$ is optimized for the point spread
function (PSF) of the instrument. The PSF is characterized by the scale $\theta_\mathrm{PSF}=0.1^\circ$ here. For SNRs that have to be considered
as extended on the scale of the PSF, the integration radius is chosen
such that the emission of the SNR can be assumed to be fully contained
and spillout from PSF effects is negligible. The effective area curves
were calculated separately for the two cases.
The flux upper limits are calculated at the 95\% confidence level and given in units of the flux of the Crab nebula of $\Phi_\mathrm{Crab}(E>\unit[1]{TeV})=2.2\cdot 10^{-11}\mathrm{cm}^{-2}\mathrm{s}^{-1}$.

\section{Discussion}
\label{sec:paramUL}
Supernova remnants are the most plausible sites of cosmic-ray
acceleration. As a consequence, they are predicted to be emitters of VHE
$\gamma$ radiation by $\pi^0$ decay during the early stage of their
evolution. The good sensitivity of H.E.S.S.~and the large dataset
accumulated in its Galactic Plane Survey now allow us to compare
quantitative estimates for this emission to observations, and to study
the SNR population in the Inner Galaxy as a whole. From Green's
catalog of radio SNRs, candidates were selected whose estimated age
makes them predicted sources of VHE \g\ radiation and the
existance of a distance measurement allows for the expected flux to be
calculated.

\begin{figure*}[th]
  \centering
  \includegraphics[width=5.3in,height=3in]{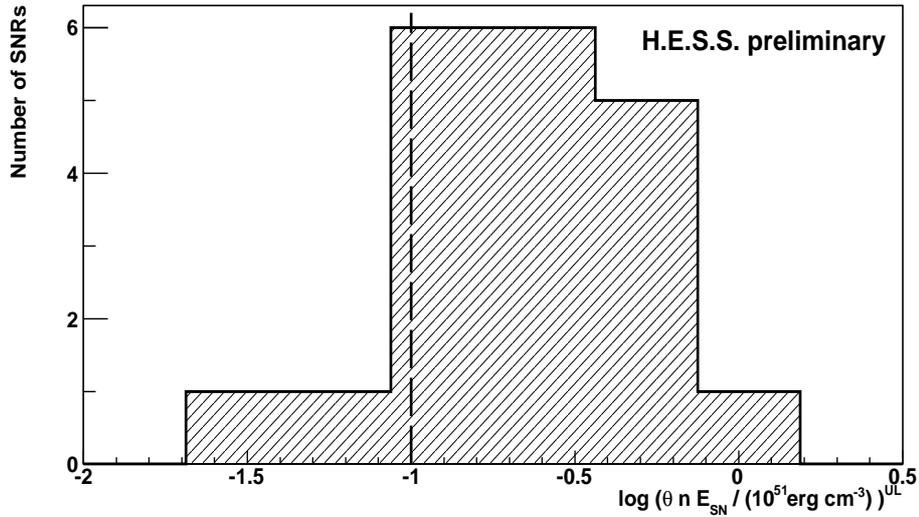}
  \caption{Distribution of the preliminary upper limit of $\theta \cdot E_\mathrm{SN} \cdot n$ for SNRs considered in the present study.}
  \label{log_p_dist}
 \end{figure*}

Fig.~\ref{log_p_dist} shows the distribution of the upper limits on
the product $p\equiv \theta{}E_\mathrm{SN}n/\unit[10^{51}]{erg}\,\mathrm{cm}^{-3}$ as calculated from the
H.E.S.S.~flux upper limit in the framework of the model of Drury \etal. The vertical line marks a value of $p=0.1$
which would be expected for a supernova explosion energy of
$E_\mathrm{SN}=10^{51}\,\mathrm{erg}$, an ambient number density of
$n=1\,\mathrm{cm}^{-3}$, and an acceleration efficiency of
$\theta=0.1$. The H.E.S.S.~upper limits fall within the same order of
magnitude of this value.

SNRs with an upper limit on the $p$ parameter above 0.1 can be divided into two groups: those with low exposure and those with detection in VHE \g\ rays. In the first case the H.E.S.S. coverage of the region is not deep enough to put a constraining upper limit on the SNR, whereas in the latter case VHE \gr\ emission has been detected, which can originate from the shell itself (e.g. G347.3-0.5, HESS~J1713-397), from a possible bright pulsar wind nebulae (PWN, e.g. G0.9+0.1, HESS~J1747-281) or can result from an overlapping bright source (e.g. G27.4+0.0, HESS~J1841-055).

There are a few SNRs that appear to be underluminous, and they are
interesting because, in principle, they have the potential of
falsifying the model of hadronic \gr\ production. But the
uncertainties in the input parameters to the model are large. In
particular, the ambient density is usually uncertain by an order of
magnitude or more and the acceleration efficiency is also not well
known.

In summary, the present study constitutes a first falsification test for a
commonly used model of the \gr\ visibility of supernova
remnants. Further observations and future
instruments like the Cherenkov Telescope Array (CTA) with its
vastly increased sensitivity and good survey capabilities will contribute to a better understanding of
acceleration mechanisms of cosmic rays.

{\small \subsubsection*{Acknowledgements}
The support of the Namibian authorities and of the University
of Namibia in facilitating the construction and operation
of H.E.S.S. is gratefully acknowledged, as is the
support by the German Ministry for Education and Research
(BMBF), the Max Planck Society, the French Ministry
for Research, the CNRS-IN2P3 and the Astroparticle
Interdisciplinary Programme of the CNRS, the U.K. Science
and Technology Facilities Council (STFC), the IPNP
of the Charles University, the Polish Ministry of Science
and Higher Education, the South African Department of
Science and Technology and National Research Foundation,
and by the University of Namibia. We appreciate
the excellent work of the technical support staff in Berlin,
Durham, Hamburg, Heidelberg, Palaiseau, Paris, Saclay,
and in Namibia in the construction and operation of the
equipment.}

\begin{table*}
\begin{center}
\begin{tabular}{|l|l|l|r|r|c|c|c|}
\hline 
  \multicolumn{1}{|c|}{SNR} &
  \multicolumn{1}{|c|}{H.E.S.S. name} &
  \multicolumn{1}{c|}{$R$} &
  \multicolumn{1}{c|}{$d$} &
  \multicolumn{1}{c|}{Age} &
  \multicolumn{1}{c|}{$\Phi^\mathrm{UL}$} &
  \multicolumn{1}{c|}{$(\theta E_\mathrm{SN} n)^\mathrm{UL}$} &
  \multicolumn{1}{c|}{Reference} \\

  \multicolumn{1}{|c|}{} &
  \multicolumn{1}{|c|}{} &
  \multicolumn{1}{c|}{{\small (deg)}} &
  \multicolumn{1}{c|}{{\small (kpc)}} &
  \multicolumn{1}{c|}{{\small (yr)}} &
  \multicolumn{1}{c|}{{\small (\% CU)}} &
  \multicolumn{1}{c|}{{\small ($10^{51} \mathrm{erg}$ cm$^{-3}$)}} &
  \multicolumn{1}{c|}{} \\
\hline
  G0.9+0.1 & HESS J1747-281 & 0.067 & 10.5 & 1900 & 4.6 & 1.24& \cite{Fang:G0009+001,Aharonian:G0009+001} \\
  G11.2-0.3 &  & 0.033 & 5.0 & 1625 & 1.6 & 0.10 & \cite{Tam:G0112-003} \\
  G12.8-0.0 & HESS J1813-178 & 0.025 & 4.7 & 1200 & 12.4 & 0.67 & \cite{Fang:G0128-000,Messineo:G0128-000,Aharonian:G0128-000}\\
  G15.9+0.2 & & 0.049 & 8.5 & 2000 & 0.2 & 0.04 & \cite{Reynolds:G0159+002,Lopez:G0159+002} \\
  G16.7+0.1 & & 0.033 & 10.0 & 5000 & 1.0 & 0.25 & \cite{Helfand:G0167+001,Hewitt:G0167+001} \\
  G21.5-0.9 & HESS J1833-105 & 0.033 & 4.7 & 870 & 3.3 & 0.18 & \cite{Bietenholz:SNRdistance,Camilo:G0215-009}, \\
  & & & & & & & \cite{Tian:G0215-009,Matheson:G0215-009,Djannati:G0215-009} \\
 G27.4+0.0 & & 0.033 & 8.5 & 1000 & 1.5 & 0.26 & \cite{Tian:SNRdistance,Vink:G0274+000} \\
  G28.6-0.1 & & 0.090 & 7.0 & 1350 & 3.0 & 0.36 & \cite{Bamba:G0286-001,Ueno:G0286-001} \\
  G29.7-0.3 & HESS J1846-029, Kes 75 & 0.025 & 6.0 & 723 & 2.8 & 0.25 & \cite{Djannati:G0215-009,Livingstone:G0297-003,Leahy:G0297-003} \\
  G31.9+0.0 & & 0.049 & 8.0 & 4050 & 0.8 & 0.13 & \cite{Chen:G0319+000} \\
  G43.3-0.2 & HESS J1911+090, W 49B & 0.029 & 10.0 & 2500 & 0.7 & 0.18 & \cite{Hwang:G0433-002,Brogan:G0433-002,Brun:G0433-002} \\
  G291.0-0.1 & & 0.116 & 3.5 & 2500 & 1.5 & 0.05 & \cite{Harrus:G2910-001} \\
  G292.0+1.8 & & 0.082 & 6.2 & 2990 & 1.6 & 0.15 & \cite{Gaensler:G2920+018,Winkler:G2920+018} \\
  G292.2-0.5 & HESS J1119-614 & 0.144 & 6.5 & 1610 & 4.2 & 0.44 & \cite{Gonzalez:G2922-005,Djannati:G2922-005} \\
  G309.2-0.6 & & 0.112 & 5.4 & 2000 & 1.7 & 0.12 & \cite{Gaensler:G3092-006,Rakowski:G3092-006} \\
  G315.4-2.3 & HESS J1442-624, RCW 86 & 0.350 & 2.8 & 1825 & 10.2 & 0.20 & \cite{Aharonian:RCW86,Caprioli:SNRs} \\
  G347.3-0.5 & RX J1713.7-3946 & 0.498 & 1.0 & 1600 & 75.7 & 0.19 & \cite{HESS:1713I,Cassam-Chenai:RXJ1713,Fukui:G3473-005} \\
  G348.5+0.1 & HESS J1714-385, CTB 37A & 0.125 & 9.0 & 1617 & 3.5 & 0.69 & \cite{Aharonian:CTB37,Caprioli:SNRs} \\
  G348.7+0.3 & HESS J1713-381, CTB 37B & 0.142 & 9.0 & 3000 & 3.5 & 0.70 & \cite{Aharonian:CTB37,Caprioli:SNRs} \\
  G349.7+0.2 & & 0.019 & 13.8 & 2800 & 1.2 & 0.55 & \cite{Castro:G3497+002,Caprioli:SNRs} \\
\hline

\end{tabular}
\caption{Parameters used for the SNRs considered within the present study, together with the preliminary values of the H.E.S.S. integral flux upper limit (at the 95\% confidence level in units of the flux of the Crab nebula) and the derived upper limit on $\theta \cdot E_\mathrm{SN} \cdot n$. The H.E.S.S. identifier is given for those sources where the SNR or an associated PWN is the most likely counterpart.}
\label{table_SNRs}
\end{center}
\end{table*}

\end{document}